# Compositional Trends in Surface Enhanced Diffusion in Lead Silicate Glasses

Ajay Annamareddy,[a,*] Manel Molina-Ruiz,[b] Donez Horton-Bailey,[b] Frances Hellman,[b,c] Yuhui Li,[d] Lian Yu,[d,e] and Dane Morgan[a,*]

[a]Department of Materials Science and Engineering, University of Wisconsin-Madison, Madison, Wisconsin 53706, USA
[b]Department of Physics, University of California Berkeley, Berkeley, California 94720, USA
[c]Lawrence Berkeley National Laboratory, Berkeley, California 94720, USA
[d]School of Pharmacy, University of Wisconsin-Madison, Madison, Wisconsin 53705, USA
[e]Department of Chemistry, University of Wisconsin-Madison, Madison, Wisconsin 53706, USA

* Authors to whom correspondence should be addressed: vkannama@ncsu.edu and ddmorgan@wisc.edu

In this work, we use molecular dynamics simulations to study the enhancement of surface over bulk diffusion (surface enhanced diffusion) in $(PbO)_x(SiO_2)_{1-x}$ glasses. This work is motivated to better understand surface diffusion in glasses and its connection to fragility, and to enhance surface diffusion in silica and related glasses for greater thermodynamic stability during vapor-deposition. By adding PbO to silica, the fragility of glass increases continuously for $10\% \leq x \leq 70\%$ during experiments. The increase in fragility may correspond to an increase in surface enhanced diffusion, as fragility and surface diffusion are correlated. We observe that for the silicates investigated, while surface enhanced diffusion increases with fragility, the enhancement is quite small. The slower diffusing Si and O atoms have higher enhancements, which could allow for some surface stabilization effects. We demonstrate that there are only small changes in atomic arrangements, consistent with the similar diffusion rates, at the surface as compared to bulk. Finally, we examine the trend of bulk versus surface diffusion in view of previous observations in organic and metallic glasses and found that in oxides, fragility increase may not be strongly linked to enhanced surface diffusion.

## 1. INTRODUCTION

Surface diffusion (quantified by $D_S$) refers to the motion of atoms or molecules at the surface of a material and can be contrasted with bulk or volume diffusion (quantified by $D_V$) exhibited by atoms or molecules deep inside the material. The weak constraints imposed on the movement of surface atoms compared to bulk allow surface diffusion to be often much higher (orders of magnitude) than bulk diffusion.[1] The ratio of these mobilities ($D_S/D_V$) can be termed as the *surface enhanced diffusion*. While the mobility of atoms on the surface of a crystal is dominated by discrete hops of a small number of lower coordinated atoms,[2] in glasses surface diffusion is attributed to the high mobility of all atoms at the surface. Very high surface enhanced diffusion is possible in glasses, and values as high as $10^8$ have been observed.[3] This high surface mobility has many consequences. In molecular glasses, rapid crystallization has been observed at the surface while remaining very slow within the bulk of the sample.[4] Under suitable deposition rates and substrate temperatures, high surface diffusion leads to efficient equilibration of incoming atoms during vapor deposition of glasses to create very low enthalpy, ultra-stable glasses.[5,6]

Greater stabilization is of interest in many glassy systems and in particular for amorphous oxides as stabilization may reduce low-energy excitations that are detrimental in quantum computing and some optical coatings applications.[7] These excitations occur in the amorphous oxide layer formed on the outside of quantum computing devices when exposed to air and introduce noise and decoherence that limit the devices' usability. However, oxides generally have rather modest surface enhanced diffusion, which can be attributed to the small fraction of broken bonds at the surface.[8] In particular, in amorphous silica ($SiO_2$), a material of wide use and particular interest for quantum computing, bulk and surface diffusion are very similar in the high temperature region investigated.[8] There is therefore significant interest in exploring ways one might increase surface enhanced diffusion in silica or related oxides, to achieve greater thermodynamic stability during vapor-deposition and hence a better-quality film for various applications.

One approach to guide the search for higher $D_S$ in silica alloys is to consider the relationship between the fragility index and surface mobility.[9] The fragility of a glass[10] qualitatively indicates the ease of mobility excitation with temperature around the glass transition. *Strong* glass formers like silica form strong and highly directional covalent bonds between neighboring atoms and the stable short and intermediate range order these materials possess degrades very little at the transition.[8] In contrast, the atoms or molecules in *fragile* glass formers interact through ionic or van der Waals forces and a sizeable degradation in their structure occurs upon heating above the glass transition. Recently, Chen *et al.* report that more fragile glasses have higher surface enhanced diffusion.[9] This result suggests that one might engineer greater surface enhanced diffusion in silica by alloying with a second component that increases fragility. In particular, the fragility of silica can be increased by adding lead oxide.[11] Pure silica is a rigid network glass former with $SiO_4$ tetrahedra connected to each other by oxygen at the corners, and the addition of lead oxide creates non-bridging oxygen atoms that disrupts the network. Lead silicate $((PbO)_x\text{-}(SiO_2)_{1-x})$ glasses can be formed with silica content as low as 30% and the glass fragility increases[11] with increasing concentration of lead for $10\% \leq x \leq 70\%$, where $x$ is the molar percentage of PbO. For large $x$, the breakdown in the $SiO_4$ linkage is compensated by the $PbO_3$ and/or $PbO_4$ structural units acting as network formers and aiding in the formation of a glass.[12,13] While the addition of lead oxide clearly changes the bonding structure and fragility, its impact on surface enhanced diffusion is not known. Molecular dynamics simulations are an efficient way to explore the surface-enhanced diffusion of lead silicates as they can be performed significantly faster and at a lower cost than experiments, although their limited time scales make the results only a qualitative guide. The goal of this work is to use molecular dynamics simulations to explore surface diffusion in lead silicates, determine whether adding lead oxide to silica is likely to result in significantly higher surface enhanced diffusion, and examine the trends between fragility and surface enhanced diffusion of the glass.

## 2. SIMULATION METHODS

Classical molecular dynamics (MD) simulations employing a two-body interatomic potential[14] are used to study lead silicates in this work. The potential consists of four terms modelling the steric repulsion of the ions due to size effects, Coulomb interactions owing to charge transfer between the ions, charge-induced dipole attractions arising from the electronic polarizability of ions, and the van der Waals attraction. The chosen interatomic potential was originally parameterized for $PbSiO_3$ (i.e., $x$=50% in $(PbO)_x(SiO_2)_{1-x}$) to generate the correct energy and length scales at the experimental density and zero pressure.[14] The potential reproduces most features of the neutron static-structure factor as well as vibrational density of states obtained using Raman spectroscopy for $PbSiO_3$.[14] Here, we apply the interatomic potential to study three lead silicate compositions, with $x$=30%, 50% and 70%. We will term these compositions as $LS_{30}$, $LS_{50}$ and $LS_{70}$. In the Results section, we show that the potential is capable of reproducing many features of these materials.

To generate a glass, the lead silicate samples were initially equilibrated in the liquid state at 2000 K for 300 ps. Subsequently, they were quenched to 300 K at the rate of $10^{11}$ K/ps in NPT (constant number of atoms (N), pressure (P) and temperature (T)) conditions with zero nominal pressure. Periodic boundary conditions (PBCs) were employed and the time step used was 1 fs. All our simulations contained 10985 ions and were executed using LAMMPS.[15] To measure bulk properties, the final configuration at the temperature of interest during quenching was used as the starting configuration for production runs under NVT (constant number of atoms (N), volume (V) and temperature (T)) conditions with PBCs in all three directions. To measure surface properties, using the same configuration obtained from the quenching process as before, free surfaces were created by extending the simulation cell boundaries by 30 Å (creating a 30 Å vacuum layer) along the $\pm$z-axis. Again, NVT conditions and PBCs (although the atoms cannot interact through the boundaries in the z-direction) were applied, and the system was initially equilibrated for 1 ns to make sure the newly created surfaces at the two edges were relaxed before the production phase begins. Atoms in the outer 5 Å along the ±z-axis were used to evaluate the properties associated with the surface.

The diffusion coefficient, $D$, is calculated based on the relationship between $D$ to the mean-squared displacement (MSD) of atoms as given by Einstein's equation:

$$D = \lim_{t \to \infty} \frac{1}{(2d)Nt} \left\langle \sum_{i=1}^{N} \left| \vec{r}_i(t) - \vec{r}_i(0) \right|^2 \right\rangle, \tag{1}$$

where $d$ is the number of dimensions, $N$ is the number of atoms involved in the summation, $\vec{r}_i(t)$ is the position vector of atom $i$ at time $t$, and $\langle \cdots \rangle$ refers to an average over different configurations (known as the ensemble average). Diffusion of a particular species $I$ ($D_I$) is calculated by restricting the summation in Eq. 1 to atoms of type $I$. For the bulk diffusion coefficient ($D_V$), $d = 3$. For surface diffusion ($D_S$), only the lateral displacement of the particles in the xy plane is considered with $d = 2$, and all atoms belonging to a layer at time $t = 0$ contribute to the diffusion of that layer at all future times. The alpha-relaxation time, $\tau$, of the bulk is extracted from the self-intermediate scattering function, defined as

$$F_s(\vec{k}, t) = (1/N) \left\langle \sum_{i=1}^{N} e^{i\vec{k} \cdot [\vec{r}_i(t) - \vec{r}_i(0)]} \right\rangle,$$

and $\tau$ is determined from $F_s(\vec{k}, \tau) = 1/e$. The wave vector $\vec{k} = 2\pi \vec{q}/L$ where $\vec{q}$ is a vector of integers and $L$ is the length of the simulation box. The magnitude of the chosen wave vector $k$ to determine $\tau$ coincides with the first maximum of the static structure factor.

## 3. RESULTS

3.1 Glass transition temperature for the 3 compositions:

To determine the glass transition temperatures $T_g$, we show in Fig. 1 the temperature variation of simulation box volume during quenching for the three compositions $LS_{30}$, $LS_{50}$ and $LS_{70}$. $T_g$ is identified as the temperature at which the rate of volume change slows down during cooling. A linear fit is used in the high temperature range 1200 K – 1500 K and the low temperature range 300 K – 600 K for all the compositions and $T_g$ is determined by the crossing of the fit lines. The glass transition temperature decreases as the concentration of PbO increases, in agreement with experiments.[11] Table 1 shows the comparison of $T_g$ obtained from MD simulation and experiments. The simulation $T_g$ values are higher for every composition and we attribute these discrepancies mostly to the very high cooling rates employed in our simulations. The relatively constant shift in temperature between MD and experiment of 133 ± 15 K over the experimental range of 175 K suggests that the MD captures glass formation physics similar to the experiments.

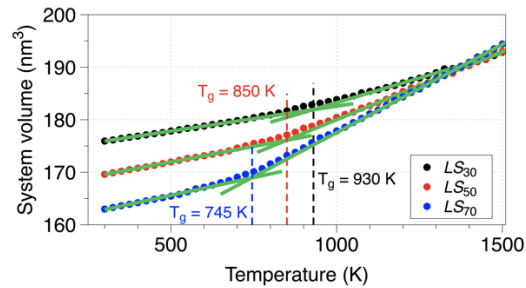

Fig. 1. Temperature variation of simulation box volume during quenching of $LS_{30}$, $LS_{50}$ and $LS_{70}$ at a rate of $10^{11}$ K/ps. Here, $LS_x$ refers to the composition $(PbO)_{x\%}(SiO_2)_{1-x\%}$.

Table 1: Comparison of MD and experimental $T_g$ for different lead silicates. The experimental $T_g$ is obtained from calorimetry measurements.[11] $\Delta T_g$ shows the difference in $T_g$ from MD and experiment.

| Composition | MD $T_g$ (K) | Expt. $T_g$ (K) | $\Delta T_g$ (K) |
|---|---|---|---|
| $LS_{30}$ | 930 | 800 | 130 |
| $LS_{50}$ | 850 | 700 | 150 |
| $LS_{70}$ | 745 | 625 | 120 |

3.2 Fragility of lead silicates:

The fragility of $(PbO)_x(SiO_2)_{1-x}$ has been observed experimentally to increase with increasing concentration of lead oxide for $10\% \leq x \leq 70\%$.[11] The fragility increase can be attributed to a disruption of the $SiO_4$ network connectivity that softens the overall structure. Here, we used simulations to study the fragility of lead silicates to demonstrate that our potentials reproduce the experimental trends. Fragility is a measure of the slowdown of the dynamics during cooling towards the glass transition temperature. It is quantified using the fragility index, $m$ defined as:

$$m = \left.\frac{\partial \log \eta(T)}{\partial (T_g/T)}\right|_{T=T_g} \cong \left.\frac{\partial \log \tau(T)}{\partial (T_g/T)}\right|_{T=T_g}. \qquad (2)$$

$m$ can be assessed from the measurements of viscosity ($\eta$) or viscoelastic relaxation time ($\tau$). Fig. 2(a) shows the experimental viscosity values of lead silicates at different temperatures, along with our estimation of $m$ for these glasses.

With the limited time scales associated with MD simulations, a direct calculation of the viscosity using the standard Green–Kubo method[16] is beyond reach for temperatures close to $T_g$. However, $m$ can also be derived from the alpha relaxation time ($\tau$) which is accessible from MD simulations by utilizing the Kohlrausch–Williams–Watts (KWW) function, $f(t) = A \exp[-(t/\tau)^\gamma]$, with A and $\gamma$ as fitting parameters, to obtain a fit to $F_s$ that can be extrapolated for long times. Fig. 2(b) shows that while $F_s$ did not decay to $1/e$ during the simulation, the KWW fit can be used to estimate $\tau$ at which $F_s$ equals $1/e$. $m$ can be determined from $\tau$ by using Eq. 2 but due to our relatively coarse temperature mesh and the fluctuations that occur in the MD simulations, it is very difficult to accurately determine the desired derivative at $T_g$. Therefore, as explained in Ref.,[17] in simulations one often uses the ratio $m^* = \tau(T_g)/\tau(1.25T_g)$ as a measure of fragility, with higher $m^*$ indicating faster slowdown with cooling towards the glass transition and hence more fragile glasses. While $m^*$ is useful to discern the fragility trends, it is not equal to the experimental fragility, and there is no known quantitative relationship between $m^*$ and $m$. As shown in Fig. 2(a), $m^*$ increases with increasing presence of lead, similarly to fragility in experiments, although $LS_{50}$ and $LS_{70}$ have very similar $m^*$. We have previously measured the fragility trends of 10 binary metallic glass formers using a similar approach and the $m^*$ values lie in the range 430 – 4900,[17] making them all more fragile than the lead silicate compositions in the current study. Hechler et al. have experimentally studied three compositions of lead silicate and have observed that all of them are stronger glass formers than any known bulk metallic glass.[18]

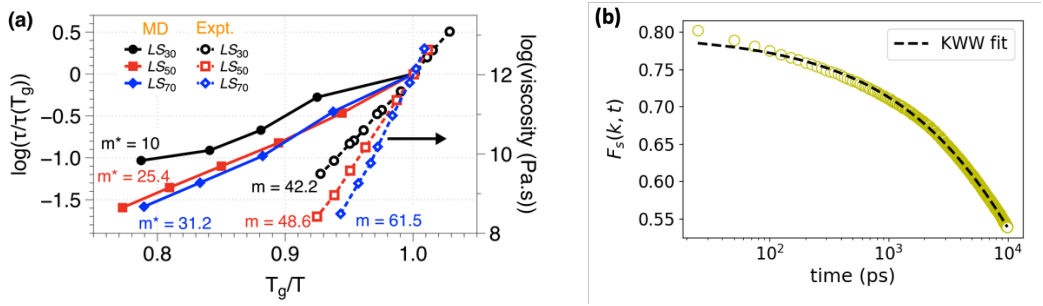

Fig. 2. (a) Variation of MD calculated $\tau(T)/\tau(T_g)$ and experimentally measured viscosity versus normalized inverse temperature for lead silicates. (b) Evolution of the intermediate scattering function $F_s(\mathbf{k},t)$ over time for $LS_{30}$ at 1000 K. The Kohlrausch–Williams–Watts (KWW) fit was used to estimate $F_s$ at longer times to obtain $\tau$.

To further confirm the trends in fragility determined by our calculation of $m^*$, we consider a structural approach to examine the change in fragility with the addition of PbO to silica. The structural approach is to analyze the connectivity between the $SiO_4$ tetrahedra. In pure silica, the basic structural unit is a $SiO_4$ tetrahedron and each oxygen atom is a bridging oxygen (BO) connecting two neighboring tetrahedra, as shown in Fig. 3(a). Nemilov had proposed a correlation of fragility with the average number of BO per polyhedron, $\bar{Q}$, and noted that the fragility increases with decreasing $\bar{Q}$.[19,20] As PbO is added to silica, each silicon atom is still bonded to 4 oxygen atoms for all compositions. However, the average number of BO associated with a silicon atom decreases with increasing lead hinting at the degradation of the network connectivity. We use $Q^n$ to denote a silicon atom having $n$ BO; $n$ can vary from 0 to 4. E.g., in pure silica, all silicon atoms are $Q^4$. Figure 3(b) shows the distribution of $Q^n$ for different compositions in lead silicates. The breakdown in the connectivity with increasing concentration of lead is quite evident as the peak fraction shifts towards lower $Q^n$ values. Neutron diffraction combined with reverse Monte Carlo have investigated the $SiO_4$ connectivity in lead silicates, $(PbO)_x(SiO_2)_{1-x}$.[13] For $x=34\%$, they demonstrated that $Q^3$ is dominant and constitutes ~ 40% of the total tetrahedra. For $x=50\%$, $Q^2$ was shown to be dominant and a symmetrical distribution centered on $Q^2$ was observed. Finally, for $x=65\%$, it was observed that $Q^0$ and $Q^1$ constitute the majority of the tetrahedra suggesting that the network connectivity is completely broken. Our MD results from Fig. 3(b) are in excellent agreement with these qualitative experimental observations for all 3 cases. Also, $\bar{Q}$ decreasing with increasing $x$ indicates a rise in the fragility of lead silicates with the concentration of lead oxide, consistent with our above calculations of $m^*$. Now that the trends in fragility in our simulations are established, we consider the trends in diffusion.

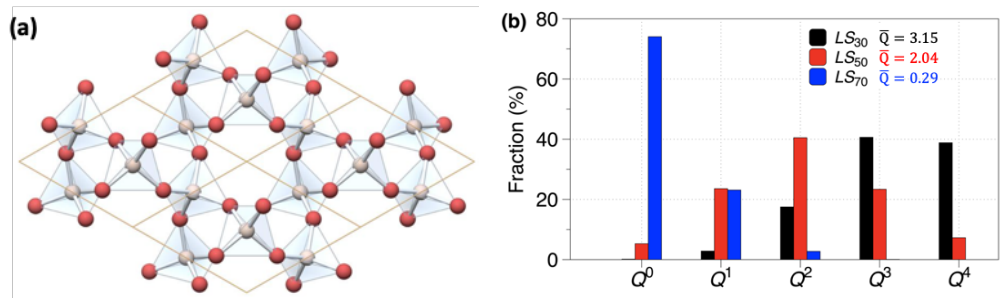

Fig. 3. (a) A simplified representation of silica $(SiO_2)$ structure.[21] The white atoms represent silicon and red atoms represent oxygen. The network structure is formed by $SiO_4$ tetrahedra connected to each other by oxygen atoms at the corners. (b) The distribution of $Q^n$ for $SiO_4$ tetrahedra for different lead silicates. The average number of BO per tetrahedron $(\bar{Q})$ is also indicated.

3.3 Bulk and surface diffusion:

Fig. 4(a) shows the temperature variation of bulk diffusion coefficient in an Arrhenius form for the three different lead silicates. The arrows point to $D_V$ at the glass transition $T_g$. For each composition, the dynamics in Fig. 4, encompassing both supercooled and glassy states, can be described by a single Arrhenius curve and no change in behavior while undergoing the transition is observed, a clear indication of a strong glass. This behavior is in contrast with the more fragile metallic glasses, where the dynamics change from Vogel-Fulcher-Tammann (VFT), a super-Arrhenius behavior[22] that diverges at a finite temperature, to Arrhenius at $T_g$ during cooling.[23] Fig. 4(b) shows the diffusion coefficients of all species separately. Pb has the highest diffusion at all compositions, while Si has the lowest diffusion, although comparable to oxygen especially for $LS_{30}$. At $T_g$, $D_{V,Pb}$ is 1660, 400 and 50 times $D_{V,Si}$ for $LS_{30}$, $LS_{50}$ and $LS_{70}$, respectively. We will see later that the low surface diffusion observed for one of the compositions can be partly explained by the reduced number density of Pb atoms at the surface.

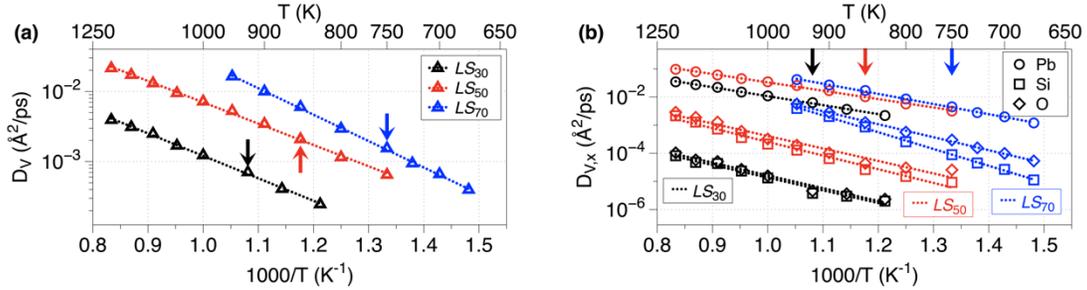

Fig. 4. (a) Arrhenius plot of total bulk diffusion coefficient ($D_V$) with temperature for different lead silicates. The dotted lines represent the best linear fit between log ($D_V$) and 1000/$T$, and the arrows represent $T_g$. (b) Plot of bulk diffusion coefficient for each component $D_{V,x}$ (x = Pb, Si and O) with temperature in Arrhenius form. The dotted lines represent the best linear fit between log ($D_{V,x}$) and 1000/$T$ and the arrows represent $T_g$. The black points and lines correspond to $LS_{30}$, red to $LS_{50}$ and blue to $LS_{70}$. Circles represent bulk diffusion coefficient for lead, squares for silicon and diamonds for oxygen.

We performed surface diffusion calculations by inserting vacuum in the ±z-direction, and the outer layers of this slab simulation, bordering vacuum, act as free surfaces. Atoms in the outer layers (shown in green in Fig. 5(a)) contribute to the calculation of surface properties, such as the surface diffusion coefficient, and all the results shown are averaged over the two outer layers in the ±z-direction. By ignoring the inserted vacuum layer, the surface layer is generally chosen as 5 Å thick unless otherwise stated. A few atoms were slightly drifted into the vacuum layer, and they were also considered as belonging to the first (surface) layer when measuring surface properties. For all compositions, oxygen atoms are found to populate the outermost monolayer. This is expected as oxygen atoms have a lower coordination number than Si or Pb, so terminating with oxygen will break fewer bonds and leads to a lower surface energy. Both experiments and simulations have demonstrated that the impact of surface on the dynamics of atoms is limited to a few nm.[24–26] In Fig. 5(b), we show that the diffusion coefficient of atoms in the middle region of the slab simulation (red atoms in Fig. 5(a)) closely resembles the bulk diffusion coefficient, indicating that our simulation box size is sufficient enough to contain the effect of surfaces to less than half the box dimensions. The deviations in Fig. 5(b) may be mainly attributed to the compositional differences at the middle of the slab simulation from the bulk composition of the glass former.

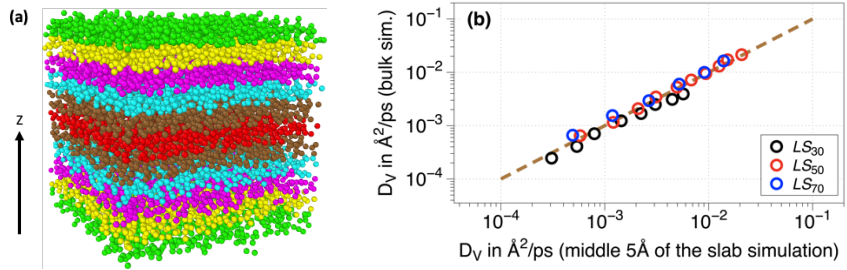

Fig. 5. (a) Depiction of a slab simulation, with atoms from different layers (each of 5 Å width) shown in different colors. The open source software OVITO[27] is used for visualization. (b) Comparison of the bulk diffusion coefficient obtained from atoms belonging to the middle region of the slab simulation (red atoms in Fig. 5(a)) and from the bulk simulation, for lead silicates at different temperatures.

We next calculate the diffusion coefficient of different layers depicted in Fig. 5(a), and, in Fig. 6(a), the depth variation of diffusion normalized to the diffusion at the middle of the slab (= $D_V$) is shown for different compositions at their respective $T_g$. Surface diffusion corresponds to the diffusion of the outermost layer. Although surface diffusion is generally expected to be higher than bulk diffusion, a marked contrasting behavior is observed for $LS_{30}$. In the next section, by pointing out the similarities in the structure and bonding at the bulk and surface, we propose that similar values of bulk and surface diffusion are not unreasonable. The lower surface diffusion as compared to bulk diffusion for $LS_{30}$ will be discussed below.

We first note that $D_{Pb}$ of different layers exhibits a very similar behavior to the total diffusion coefficient $D$ (open and solid symbols in Fig. 6(a), respectively). As $D_{Pb}$ is much greater than either $D_{Si}$ or $D_O$ (Fig. 4(b)), the behavior of lower $D_S$ compared to $D_V$ for $LS_{30}$ can mostly be understood in terms of lower $D_{S,Pb}$ compared to $D_{V,Pb}$. The lower $D_{S,Pb}$ can be attributed to the smaller number density of Pb atoms at the surface. Fig. 6(b) shows the planar number density (number of atoms per unit area) of Pb atoms normalized to the average planar number density, along the z-direction. The shaded area in Fig. 6(b) represents the width of a free surface layer chosen in this study, and a low density of Pb atoms in this region can be observed. The change in Pb number density is a consequence of the segregation of oxygen atoms to the surface. For $LS_{30}$, we estimate the fraction of PbO at the surface (outer 5 Å) to be ~ 24.5%. We propose that the reduced concentration of Pb at the surface causes a reduced surface diffusion relative to bulk. More quantitatively, from the bulk simulations, it can be observed that $D_{V,Pb}$ increases with the PbO concentration, $x$. A linear relation, shown in the inset of Fig. 6(b), can be used to approximate the relation $D_{V,Pb} = 0.00070 \, x + (-0.014)$, from which we can predict that $D_{V,Pb}(x=24.5\%)/D_{V,Pb}(x=30\%) \sim 0.45$. This ratio is close to the $D_{S,Pb}/D_{V,Pb} = 0.65$ observed for $LS_{30}$ from Fig. 6(a), suggesting that the major contribution of $D_{S,Pb}/D_{V,Pb}$ being less than 1 is simply the change in composition at the surface versus the bulk. Segregation of oxygen to the surface alters the surface number density of lead even for $LS_{50}$ and $LS_{70}$ although the effect is, as seen from Fig. 6(b), not as pronounced as in $LS_{30}$. It should be noted that Si and O atoms show modest surface-enhanced diffusion at all compositions, and $D_{S,Si}/D_{V,Si} \sim 5$ and $D_{S,O}/D_{V,O} \sim 4$ for $LS_{70}$ as shown in Figs. 6(c) and 6(d).

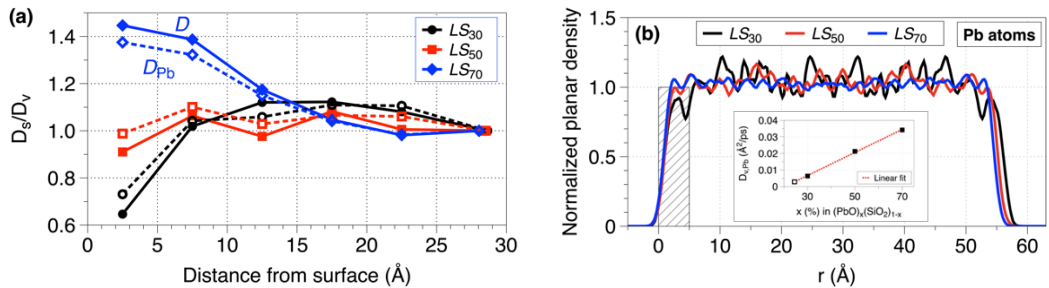

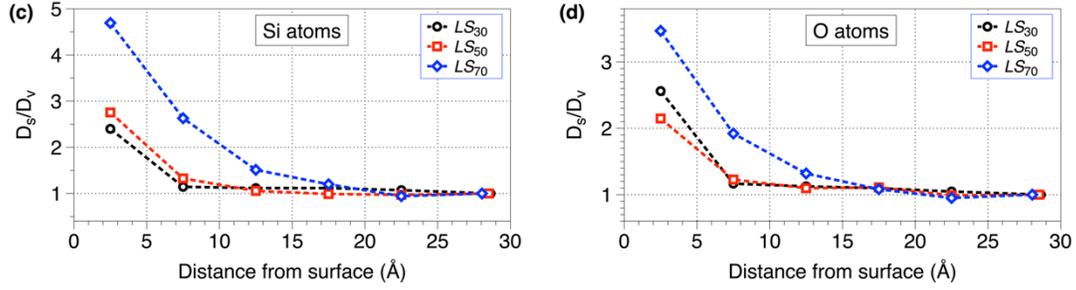

Fig. 6. (a) Total diffusion coefficient normalized to the diffusion of the central layer of the simulation box, representing bulk, versus depth in a slab simulation, shown as solid symbols. The diffusion coefficient of lead is shown as open symbols. All measurements are at $T_g$. (b) Normalized planar number density of lead versus distance from a free surface at $T_g$. The shaded region indicates the width of the outermost layer used to calculate surface properties. (inset) Variation of bulk diffusion coefficient of lead versus the composition $x$ in lead silicates at 925 K ($T_g$ of $LS_{30}$). A linear fit is used to extrapolate the diffusion at a new composition. (c) and (d) $D_S/D_V$ of silicon and oxygen atoms for different layers in a slab simulation for $LS_{30}$, $LS_{50}$ and $LS_{70}$ at $T_g$.

To examine the relation between fragility and surface enhanced diffusion ($D_S/D_V$), we can consider either the total diffusion or the diffusion of the individual species. In almost all cases, for the range 30%-70% PbO, the increase of fragility with the addition of PbO leads to an increased surface enhanced diffusion. However, as described earlier, high surface diffusion is of interest in this work as the fast mobility at the surface allows better equilibration of the as-deposited atoms to produce a more stable glass during vapor deposition. Such equilibration may be limited by the slowest atoms, in this case silicon. If this is the case, by enhancing the speed of the silicon, $LS_{70}$ glass would have more efficient equilibration compared to $LS_{30}$ and $LS_{50}$. However, this enhancement is modest, particularly in the rapidly cooled glass in the MD, and likely to have limited impact (although see estimates of surface enhanced diffusion for silicon for a more aged glass in Sec. 3.5).

3.4 Comparison of structure at the bulk and surface:

In this section, the short-range order at the bulk and surface is investigated to understand the low values of surface enhanced diffusion in lead silicates. We first focus on the structure of the $SiO_4$ tetrahedra. The spatial distribution of Si-O atoms in both bulk and surface are examined using the partial radial distribution function (rdf), $g_{Si-O}$, as shown in Fig. 7(a). In general, $g_{\alpha\beta}(r)$ gives the probability of finding an atom of type $\beta$ at a separation $r$ from an atom of type $\alpha$, normalized by the number density of $\beta$ atoms in the simulation box. In liquids and glasses, at long $r$, $g_{\alpha\beta}(r)$ equals 1. In the calculation of $g_{\alpha\beta}(r)$ for the slab simulation to study the surface atoms distribution, the extra vacuum inserted is ignored so that the number density of atoms is the same as in the bulk simulation, and $g_{\alpha\beta}(r)$ for surface atoms at long $r$ approximately equals 0.5. For the surface $g_{Si-O}$, only Si atoms in the outer 3 Å of the slab simulation are considered to closely examine the particle arrangements near the free surface. Fig. 7(a) shows that the Si-O short-range order for $LS_{50}$ is almost identical in the bulk and the surface, indicating that a free surface does not alter the distribution of Si-O bond lengths. This can be understood by realizing that the Si-O bond is very strong and hence does not depend strongly on the environment. Roder *et al.* have made a similar observation on $g_{Si-O}$ in pure silica.[8] $g_{Si-O}$ for the surface and bulk in $LS_{30}$ and $LS_{70}$ at $T_g$ are also very similar and almost indistinguishable from the $g_{Si-O}$ for $LS_{50}$ shown in Fig. 7(a). Differences observed in $g_{Si-O}$ of bulk and surface starting from the second shell indicate that, at these separations, surface atoms have smaller neighboring atoms compared to the atoms in the bulk, and at large $r$ the surface atoms have approximately half the neighbors compared to bulk atoms. A comparison of the atomic arrangements in the bulk and surface for the metallic glass former $Cu_{50}Zr_{50}$, having $D_S/D_V$ of ~ 26 at $T_g$, is also shown in Fig. 7(a). In this case, there is clearly a loss of first neighbors for the surface atoms that contributes to the higher surface enhanced diffusion.[17] The more similar local structure for Si in these simulations compared to the metal is consistent with the lower $D_S/D_V$ of ~ 2–5.

Next, we examine the connectivity of the SiO$_4$ tetrahedra, quantified by the average bonded oxygen per tetrahedron $\overline{Q}$, at the surface and compare with the connectivity in the bulk already shown in Fig. 3. In network glass formers, diffusion generally occurs on a microscopic level through the localized tear and repair of discrete covalent bonds. Hence, a smaller (larger) value of $\overline{Q}$ at the surface as compared to bulk will make the bond-breaking process easier (harder) for oxygen atoms at the surface, which will affect the diffusion of both oxygen and silicon atoms. In Fig. 7(b) we show the variation of $\overline{Q}$ versus normalized depth from a free surface. Both the ends on the abscissa of Fig. 7(b) represent free surfaces and their $\overline{Q}$ are shown as open symbols. Interestingly, $\overline{Q}$ is almost invariant with the distance from a surface and its value closely matches the $\overline{Q}$ obtained from bulk simulation, shown as a dotted line for each composition. The largest deviation of surface $\overline{Q}$ from the bulk is observed for $LS_{70}$ but $\overline{Q}$ for both surface and bulk are quite small. These results so far show that SiO$_4$ tetrahedra have similar bond lengths and connectivity at the surface and bulk. One metric related to Si-O bonding that showed a change in character from the bulk to surface is the Si-O-Si (inter-tetrahedral) bond angle distribution, shown in Fig. 7(c) for $LS_{30}$. While the main peak for the surface atoms is around $150^0$ similar to in the bulk, there is the emergence of a new peak around $90^0$. This new peak at a smaller angle indicates the closing of the Si-O-Si angle and is predominantly manifested by oxygen atoms in the outer monolayer. However, we are not certain if the closing of the Si-O-Si angle at the surface impacts the oxygen bond-breaking process. $LS_{50}$ and $LS_{70}$ also show the new peak, but it is more pronounced in $LS_{30}$. We next study how the mobile species Pb is distributed in the bulk and surface.

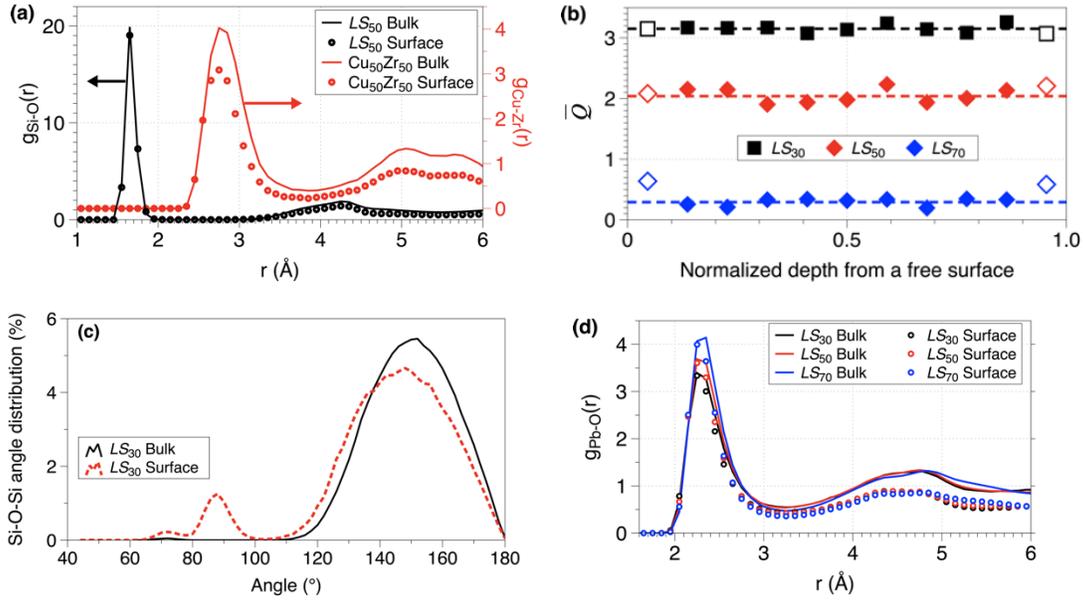

Fig. 7. (a) Partial radial distribution function of Si-O atoms in bulk (solid line) and surface (points) for $LS_{50}$ at $T_g$. A comparison to $g_{Cu\text{-}Zr}$ in Cu$_{50}$Zr$_{50}$, which has good surface enhanced diffusion as observed from MD simulations, at $T_g$ (700 K) is also shown (shifted slightly along the vertical axis for clarity). (b) Variation of the average bonding oxygen (BO) per SiO$_4$ tetrahedron versus depth from a free surface in different lead silicates. The open symbols represent $\overline{Q}$ associated with free surfaces. (c) Si-O-Si bond angle distribution for the bulk and surface in $LS_{30}$ at $T_g$. (d) Variation of $g_{Pb\text{-}O}$ in the bulk (solid line) and surface (points) for $LS_{30}$, $LS_{50}$ and $LS_{70}$ at $T_g$.

While silicon mainly bonds with four oxygen atoms at all compositions, lead atoms distribute around oxygen in many ways depending on the concentration of lead oxide. For low $x$, PbO acts as a network modifier[12,18] with lead atoms settling at interstitial positions of the rearranged SiO$_4$ network, surrounded by the non-bridging oxygen. For high concentrations, PbO acts as a network former and forms $PbO_n$ polyhedra ($n$=3–5, $n$=4 is major).[13] Based on this, lead atoms will have low oxygen coordination at low $x$ and vice-versa. Fig. 7(d) shows $g_{Pb-O}$ for different PbO concentrations for the bulk and surface. The surface $g_{Pb-O}$ is calculated by only considering Pb atoms in the outer 3 Å of the slab simulation. As expected, in the bulk, the first peak increases with increasing $x$, indicating greater coordination of oxygen atoms for lead. Comparing the bulk and surface, the short-range order (up to the first minimum of $g_{Pb-O}$) is very similar, with the surface having a slightly lower coordination than the bulk. Compared with the bulk, the oxygen coordination for lead atoms at the surface is 11%, 12% and 15% lower for $LS_{30}$, $LS_{50}$ and $LS_{70}$. Hence, the lead atoms at the surface are similarly enveloped by oxygen atoms just as in the bulk. For comparison, the loss of first neighbors in Fig. 7(a) for Cu atoms in $Cu_{50}Zr_{50}$ is 30%.

Overall, the surface and bulk coordination shells are found to be similar for the Pb and Si cations, consistent with the values of $D_S/D_V$ being fairly close to one. However, the coordination shells decrease noticeably more for Pb than for Si, while $D_S/D_V$ is higher for Si than for Pb. This suggests that the reduced first neighbor coordination shell is not a robust guide for modest trends in $D_S/D_V$. This limitation is not surprising, as the detailed diffusion mechanisms are likely sensitive to other factors than just first-neighbor coordination, e.g., local composition, second-neighbor shells, and bond angles.

3.5. Correlation of fragility and surface-enhanced diffusion:

In this section, we discuss how the predicted bulk and surface diffusion from this work can be extrapolated from MD cooling rates to experimentally relevant cooling rates, and how the extrapolated values are related to the earlier observed experimental trends of diffusion with fragility.[9,17,28] In order to compare our results to those analyzed from experiments, it is necessary to estimate $D$ values for a glassy state similar to those measured in experiments, which are far more relaxed than those obtained from our MD quench process. This estimation can be done by extrapolation using the approach developed by Chen et al. in Ref.[9] that demonstrated (Fig. 5 therein) that different glasses have a linear relation between experimental log $D_S$ and log $D_V$. In our previous work,[17] we showed that this linear trend between log $D_S$ and log $D_V$ is also present in the simulated metallic glass data. This linear trend can be used to estimate the experimental surface enhanced diffusion $D_S/D_V$ by extrapolation. We assumed a fixed $D_V$ value of the glass former of $10^{-22}$ m$^2$/s at $T_g$, which is consistent with typical experimental values. Fig. 8(a) shows the fit of log $D_S$ against log $D_V$ for $LS_{70}$ extrapolated to $D_V = 10^{-22}$ m$^2$/s. The surface diffusion is barely higher than the bulk diffusion after extrapolation. Following the same procedure, for $LS_{30}$ and $LS_{50}$ the $D_S/D_V$ ratio is less than 1, and we attribute this to segregation of oxygen at the free surface that alters the composition of the surface compared to the bulk. The precise surface enhanced diffusion $D_S/D_V$ at $T_g$ from the extrapolation corresponding to $D_V = 10^{-22}$ m$^2$/s are 0.38 ($R^2$ of the fit = 0.996), 0.24 ($R^2$ = 0.998), and 1.23 ($R^2$ = 0.999) for $LS_{30}$, $LS_{50}$ and $LS_{70}$.

While pure silica ($m = 20$)[29] and $LS_{70}$ ($m = 61$) have quite different fragilities, their surface-enhanced diffusions are very similar (~ 1). This suggests that for these oxides, a higher fragility of the glass may not point to a higher surface-enhanced diffusion, unlike in organic glasses[9] and metallic glasses.[17] However, as noted above, it may be the surface-enhanced diffusion of the slowest elements that are most important for creating more stable glasses, and these are not well represented by the total diffusivity in $(PbO)_x(SiO_2)_{1-x}$ compositions with $x = 30\%-70\%$ since Pb diffusivity dominates the total diffusion. Therefore, using the extrapolation approach, we also measured the $D_S/D_V$ for all individual species. The results for $LS_{70}$ are shown in Figs. 8(b), 8(c) and 8(d), and $D_{S,Si}/D_{V,Si}$ ~ 200. This enhanced surface diffusion of ~ $10^2$ is quite modest compared to that observed or estimated for many organic[9] and metallic glasses[17] cooled at experimental rates, which typically have values of $10^5$–$10^8$. Thus our overall conclusions above for $D_S/D_V$ also apply to $D_{S,Si}/D_{V,Si}$, and the surface enhanced diffusion is not dramatically increased for $(PbO)_x(SiO_2)_{1-x}$. However, $D_{S,Si}/D_{V,Si}$ ~ 200 may be enough to help silicon atoms, and, therefore, all the atoms, rearrange efficiently leading to a more stable glass. Further studies are needed on the exact scale of this enhancement for experimentally relaxed glasses and the ability of this likely relatively modest enhancement to impact stability (e.g., through vapor deposition).

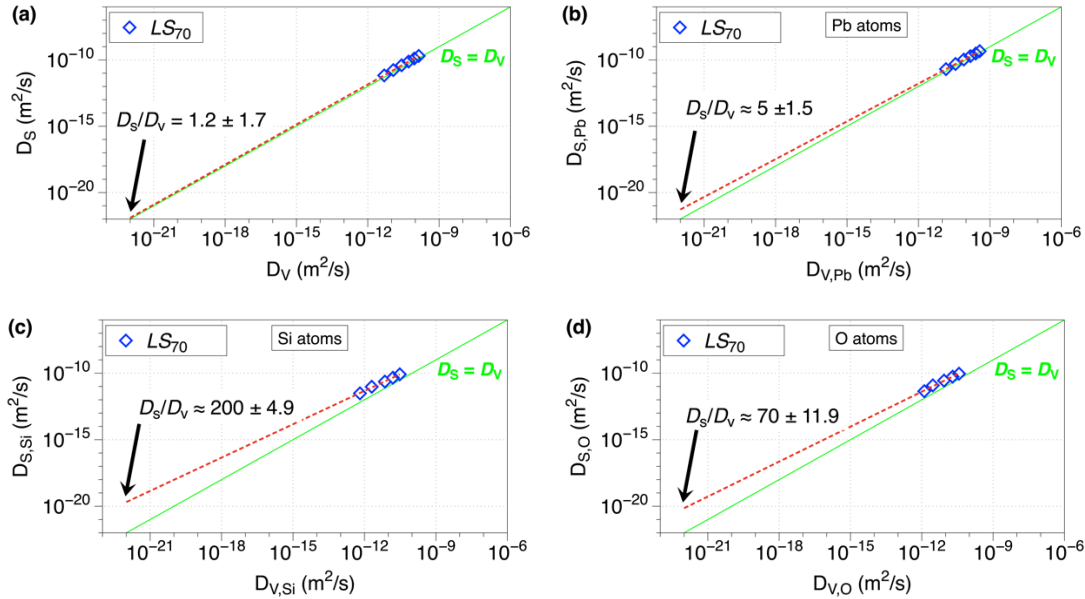

Fig. 8. (a) Plot of MD evaluated $D_S$ against $D_V$ in $LS_{70}$. The red dotted line shows the linear fit between log $D_S$ and log $D_V$. The dotted line is extended to extrapolate the $D_S$ corresponding to a typical experimental $D_V = 10^{-22}$ m$^2$/s. The extrapolation scheme is applied to the bulk and surface diffusion of (b) lead, (c) silicon and (d) oxygen atoms in $LS_{70}$. The standard deviations of the extrapolated $D_S/D_V$ are also shown, and these are derived from the standard deviations of the slope and intercept of the linear fit.

## 4. CONCLUSIONS

We studied three lead silicate $(PbO)_x(SiO_2)_{1-x}$ compositions, $x$ = 30%, 50% and 70% using molecular dynamics simulations to examine the relation between surface-enhanced diffusion and fragility. Manipulating the surface-enhanced diffusion in silica and related compositions could enable more stable glasses during vapor deposition with relevance in quantum computing applications. The fragility of the glasses was shown to increase with increasing $x$, in agreement with experiments. We observe that while surface enhanced diffusion increases with a rise in glass fragility, the enhancement is quite small. However, silicon, the slowest species, shows a larger increase in surface enhanced diffusion that may allow for more efficient equilibration at the surface leading to a more stable glass. We demonstrate that there are only small changes in atomic arrangements at the surface as compared to bulk, consistent with the similarities in diffusion rate at the surface and bulk. Finally, we examine the trend of higher surface enhanced diffusion with a rise in fragility previously proposed in organic glasses from experiments and metallic glasses simulations. Our results suggest that, for these oxides, the increase in fragility may not be strongly linked to an enhanced surface diffusion.

**Acknowledgments**

The authors are grateful to the Extreme Science and Engineering Discovery Environment (XSEDE), which is supported by National Science Foundation grant number OCI-1053575, and the National Energy Research Scientific Computing Center (NERSC), a U.S. Department of Energy Office of Science User Facility located at Lawrence Berkeley National Laboratory, operated under Contract No. DE-AC02-05CH11231. This work was supported by the University of Wisconsin Materials Research Science and Engineering Center (DMR-1720415).

**Data Availability**

The data for all the figures in the Manuscript are shared on Figshare at https://doi.org/10.6084/m9.figshare.19205148.v1. For any other relevant data, contact AA at vkannama@ncsu.edu.

**Declarations**

**Conflict of interest** There are no conflicts to declare.